# Groningen: Spatial Prediction of Rock Gas Saturation by Leveraging Augmented Well and Seismic Data with Classifier Ensembles

Dmitry Ivlev
Independent researcher

dm.ivlev@gmail.com

Abstract

This paper presents a proof of concept for spatial prediction of rock saturation probability using classifier ensemble methods on the example of the giant Groningen gas field. The stages of generating 1481 seismic field attributes and selecting 63 significant features are described. The effectiveness of the proposed method of augmentation of well and seismic data is shown, which increased the training sample by 9 times. On a test sample of 42 wells (blind well test), the results demonstrate good accuracy in predicting the ensemble of classifiers: the Matthews correlation coefficient is 0.7689, and the F1-score for the "gas reservoir" class is 0.7949. Prediction of gas reservoir thicknesses within the field and adjacent areas is made.

## 1. Introduction

One of the key aspects of successful field exploration and monitoring of reservoir development is the spatial prediction of hydrocarbon saturation of geological structures. Traditional prediction methods based on various types of elastic inversion of seismic data may be limited in conditions of a complex geological structure and insufficient coverage of the studied space with well data.

In such situations, machine learning algorithms can become an effective tool for the nonlinear, multidimensional generalization of knowledge obtained by geophysical methods in the well space to the entire territory covered by 3D seismic surveys.

The study proposes a new approach to knowledge transfer, which consists in predicting the probability of gas saturation of the territory using ensembles of classifiers trained on data from logging studies of hydrocarbon saturation along the well trajectory. Attributes of the seismic field are used as predictors.

A method for improving the quality of information synthesis by augmenting well and seismic data is described. The process of selecting the most informative features — attributes of the seismic field is presented.

The effectiveness of the proposed technological stack of methods and algorithms is demonstrated by the results of spatial prediction of gas saturation of the giant Groningen field, which is the largest natural gas field in Europe. The assessment of classification quality metrics was carried out on 42 test wells (blind well test), evenly selected throughout the study area from the learning process (Fig. 1).

The openness of the data (CC BY 4.0) on this deposit, as well as the extensive volume of structured geological and technical information, make it a unique testing ground for approaches based on machine learning algorithms.

## 2. Initial data

The data on the field was provided by the NAM field operator and distributed by EPOS-NL in the form of a project completed in Petrel 2018 [1], with a brief description of the build stages [2]. The project contains a single seismic cube in depth with a spatial resolution of amplitude values of 25 meters laterally and 8 meters vertically.

Within this area, 425 wells were identified with the results of interpretation of the logging curve according to the degree of gas saturation of the intervals. Based on the curves obtained, the intervals in the wells were assigned to one of two classes: intervals with values above the threshold value of 0.3 were assigned to class 1 (gas reservoir), and intervals with values below the threshold value were assigned to class 0 (dry).

Due to the anomalous results of the algorithms for extracting seismic attributes in the protruding northern parts of the seismic data, these areas were excluded from the study. Thus, the study area was 2221.7 $km^2$ (Fig. 1).

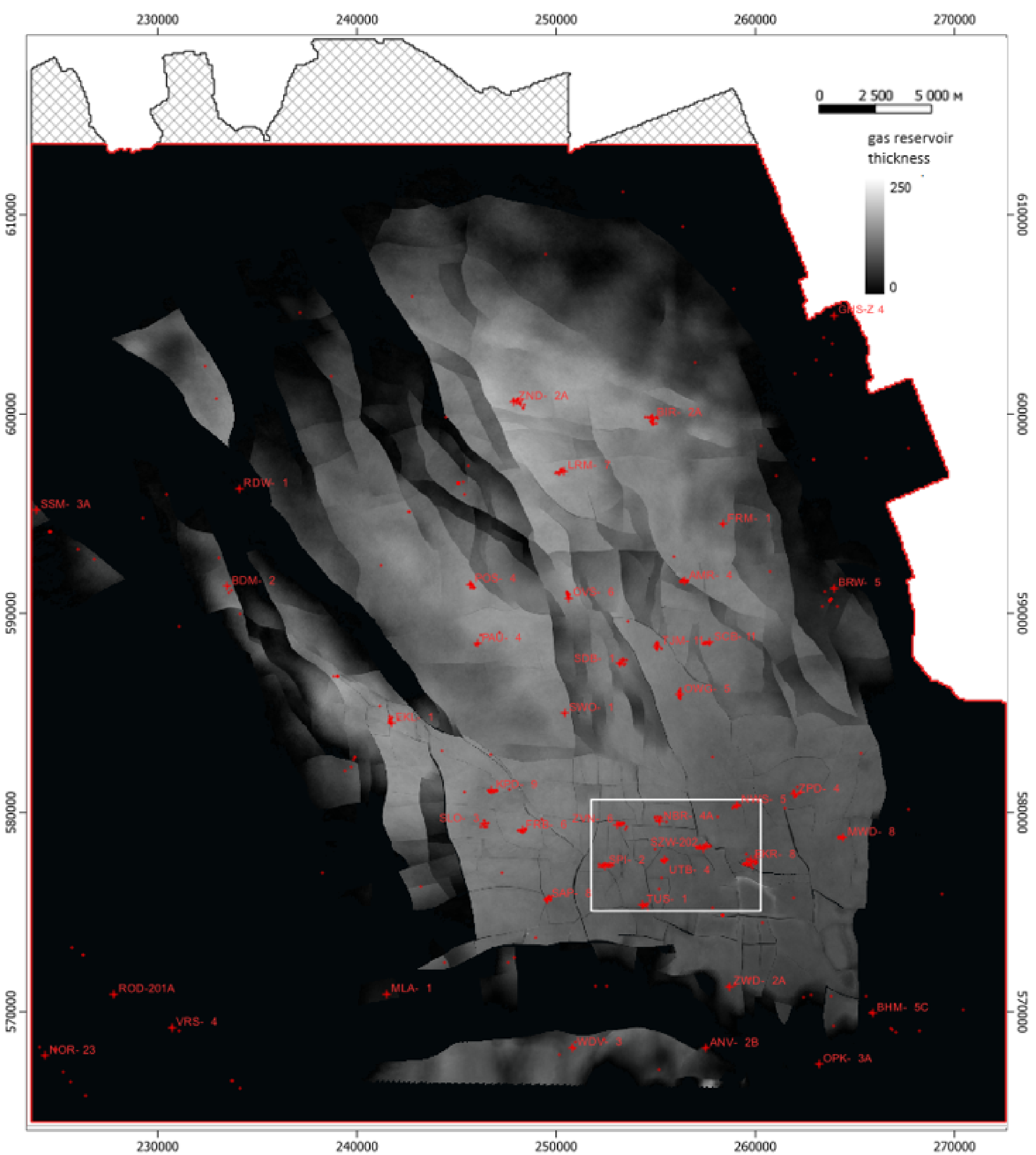

Figure 1. Prediction map of gas reservoir thickness according to the NAM model. The red line is the study area. Shaded areas - excluded parts of seismic surveys from the work. White rectangle - test area. Red dots - wells of the training dataset. Red crosses with well names - test wells. Well positions - intersection of well trajectories with the plane at 2000m

## 3. Feature engineering

The lack of computing power to process the entire array of seismic data necessitated the analysis of attributes on an area of 47.737 km$^2$, including 117 wells. This test site is representative in terms of geological structure and saturation for most of the territory, characterized by a high density of wells. It is indicated by a white rectangle in Figure 1.

In the course of the study, 31 attribute generation algorithms were applied, each of which has from zero to six configurable parameters.

A range of three to four values was set for each variable, after which combinations of these parameters were generated. As a result of the work of 31 generation algorithms, 1481 seismic attributes were obtained.

In accordance with the spatial coordinates of the well trajectory, vectors for each class label obtained from the gas saturation curve were extracted from the obtained attribute cubes (Fig. 2).

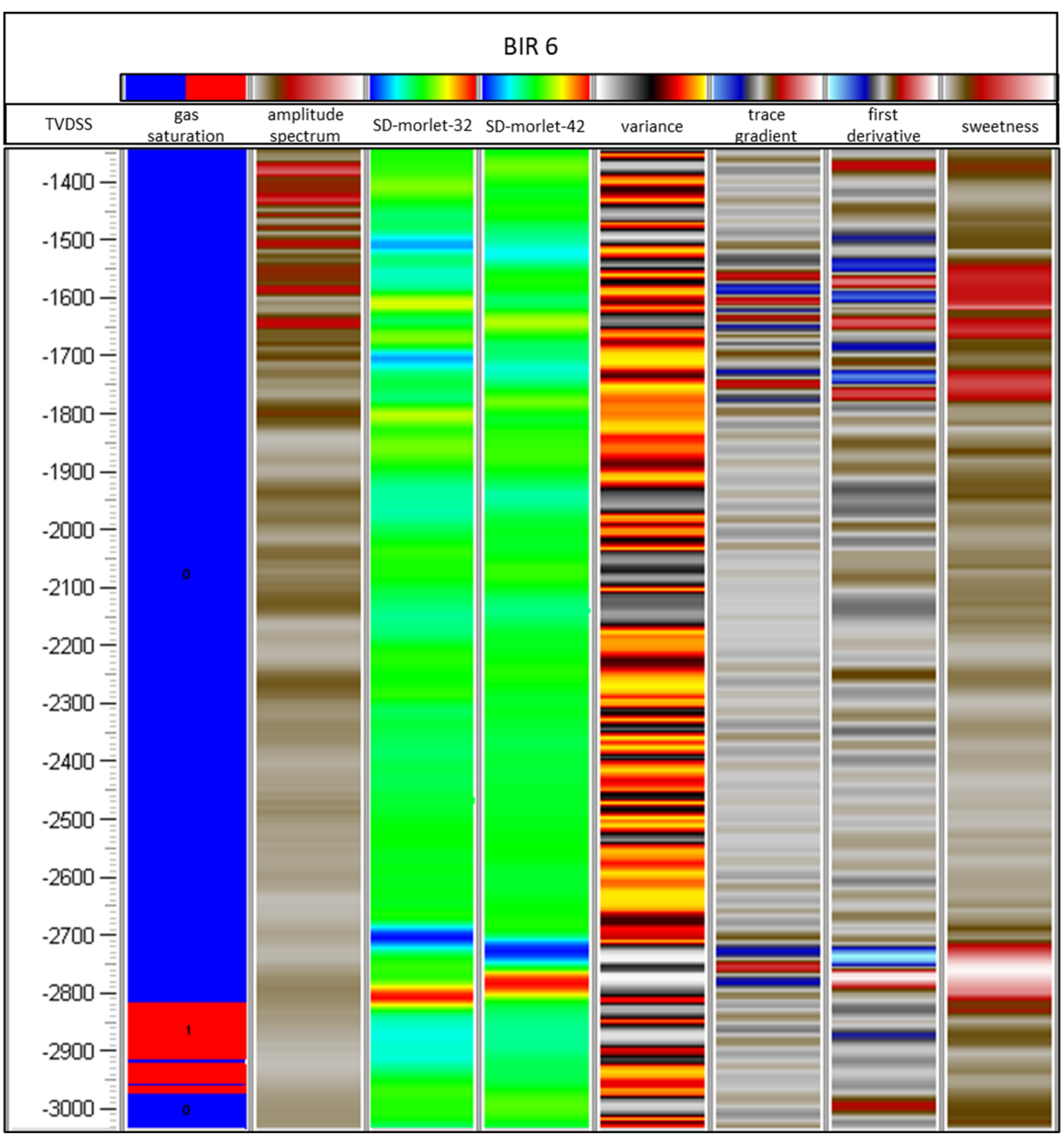

Figure 2. Example tablet with the results of saturation curve classification - class labels and a vector of 7 random seismic attributes

## 4. Feature selection

After eliminating the multicollinearity between the features, using a threshold of the coefficient of determination equal to 0.9, 716 attributes of the seismic field remained out of 1481. This means that about half of the parameters of the generation algorithms for this seismic cube did not significantly change the information about the space.

To select the most informative attributes, the method of adding noise to the training sample in the form of features with random values was used. After training the classifier and evaluating the contribution of the features to the prediction, all the features whose total contribution was less than the randomly generated ones were excluded from further consideration.

In the study, 16 features with random values were generated for each evaluation epoch, and the features were added to 716. In the next step, the classifier of gradient boosting of decision trees LightGBM [3] was trained with fixed hyperparameters, but with a different random seed initialization of the initial state for each epoch. Then, the Shapley coefficient [4] was evaluated for the trained model and a list of features with an efficiency for classification above randomly generated features was formed.

After 200 epochs, out of 716 features, only 63 were selected whose importance exceeded that of randomly generated features in all iterations. Of the 31 seismic feature generation algorithms, 15 were found to be important for prediction (Table 1).

Table 1. Attribute extraction algorithms and the number of their customization variations

| # | attribute generation algorithms | count |
|---|---|---|
| 1 | spectral decomposition | 17 |
| 2 | coherence | 10 |
| 3 | instantaneous frequency | 9 |
| 4 | variance | 6 |
| 5 | rms | 5 |
| 6 | quadrature amplitude | 3 |
| 7 | local flatness | 2 |
| 8 | structural smoothing | 2 |
| 9 | reflection intensity | 2 |
| 10 | local structural azimuth | 2 |
| 11 | sweetness | 1 |
| 12 | dtw | 1 |
| 13 | instantaneous amplitude | 1 |
| 14 | chaos | 1 |
| 15 | amplitude spectrum | 1 |
| | **sum** | **63** |

## 5. Creating dataset

### 5.1. Creating a spatial dataset

63 attributes were extracted from the total seismic data volume under study using parameters obtained from the previous attribute selection step at the test site. Due to computational limitations, the volume of rock to be predicted was reduced. Thus, the study area was covered with a 3D grid with a uniform distribution of nodes in the depth range of 2500-3500 meters, with lateral spacing of 50 meters, vertical spacing of 8 meters, and a total size of $i$ 980, $j$ 1090, $k$ 125. Seismic attribute values were assigned to the grid nodes using an averaging method. This data set was used for spatial prediction of gas saturation of the area in the inter-well space.

### 5.2. Creating a base well dataset

The values of 63 attributes for all 425 wells were extracted from seismic data in the intervals with saturation classification logging in accordance with the coordinates of their trajectories (Fig. 2). The scale of the logging curve with a sampling frequency of 20 centimeters was used for extraction. Due to the difference in the scales of seismic and logging data, a significant number of duplicate vectors encoding classes were obtained. Duplicates were deleted, when vectors collided — encoding different classes with one vector, a vector with class 0 was excluded from the training set. Thus, a base dataset was created, which contained 53 458 vectors with class labels, of which 11.62% encoded a class with gas saturation (Table 2).

### 5.3. Augmentation of well and seismic data

Data augmentation is the process of artificially increasing the volume and diversity of training data through various transformations and modifications of existing data. This method is often used in machine learning and deep neural networks to improve the quality of models and increase their resistance to overfitting.

In practice, several methods of augmentation of well and seismic data have been developed [5, 6, 7]. The methods proposed in the works use uncertainties in geological and geophysical data as a hyperparameter of the global configuration of the technological stack from machine learning algorithms through modification of the training dataset. The boundaries of the search for a reverse response are set based on geostatistics or a priori assumption about the boundaries of our knowledge about a natural object and possible instrumental errors in obtaining this knowledge.

To check the boundary conditions of potential data augmentation, a test was performed for the vertical and lateral stationarity of target classes in local areas [8]. After averaging the test results in different areas of the studied space, the p-value was obtained for the vertical component 0.0158, for the lateral component 0.0561. The null hypothesis for this test is the assumption that the series are locally stationary. Thus, at a statistical significance level of 0.05, the null hypothesis was accepted only for the lateral data series, and for the vertical, respectively, it was rejected.

An experimental semivariogram has been constructed for the lateral data series, and a theoretical exponential model has been approximated on it. The maximum distance for different directions at which the data values are correlated is determined in the range of 40-140 meters (Fig. 3).

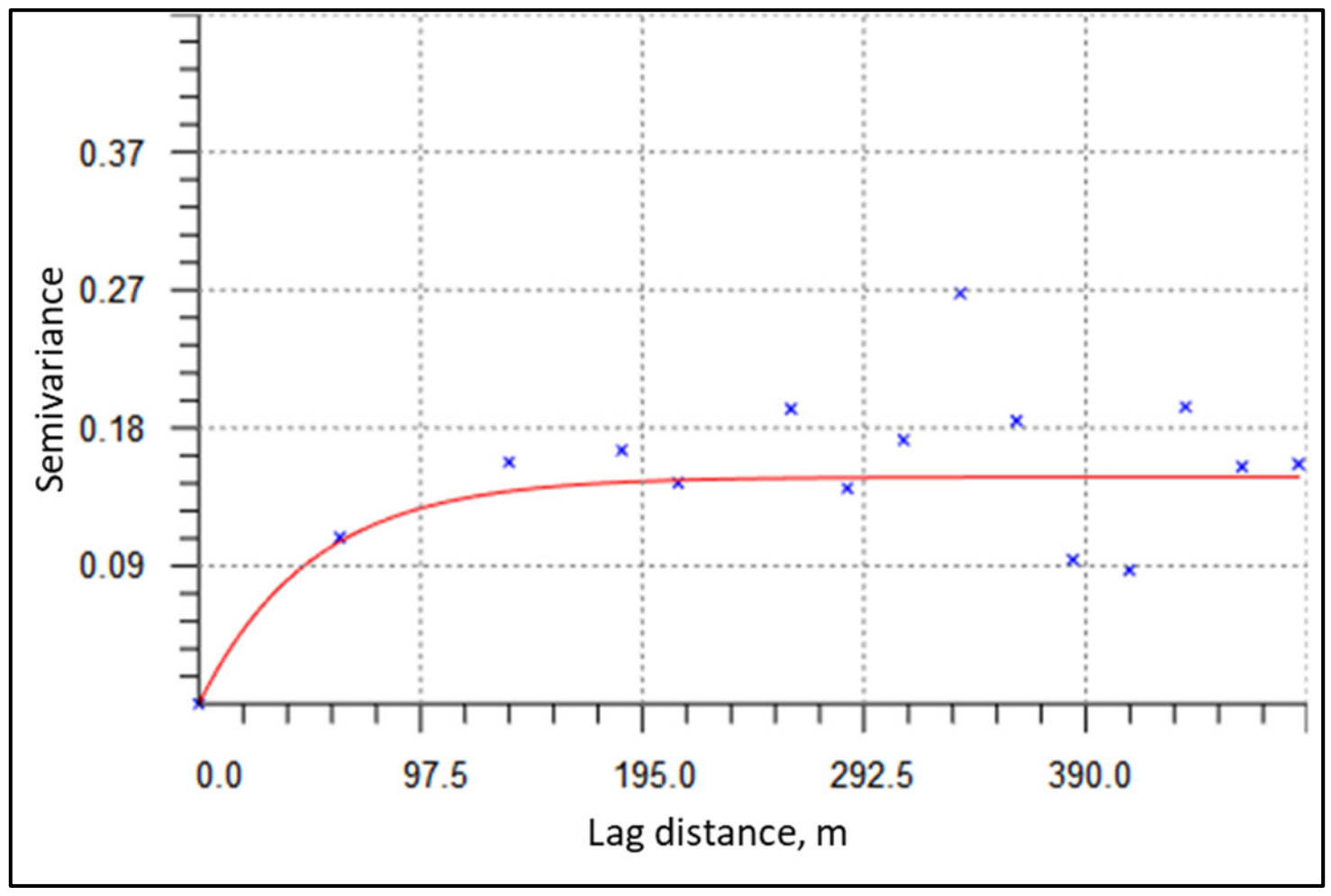

Figure 3. Experimental semivariogram of gas sands. The main direction is NW-SE. Blue dots are observed data. Red line - exponential approximation function.

Based on the results of geostatistical analysis, it can be assumed that within the lateral displacement from the well, equal to the step of seismic data (25 meters) and within 36 meters, changes in the lithology of gas-saturated sandstones are insignificant or close to the initial one.

In this regard, this paper proposes and develops an augmentation method – one step grid with zero vertical offset (sgz0). The essence of the method is to copy the trajectory of the well along with the target logging at a distance of one step (25 meters) from the initial position and one step from the new position closest to the well. This creates a square around the initial position of the well with a side of 50 meters. At the same time, a different attribute vector of the seismic field from the new position is assigned to the copied labels (Fig. 4).

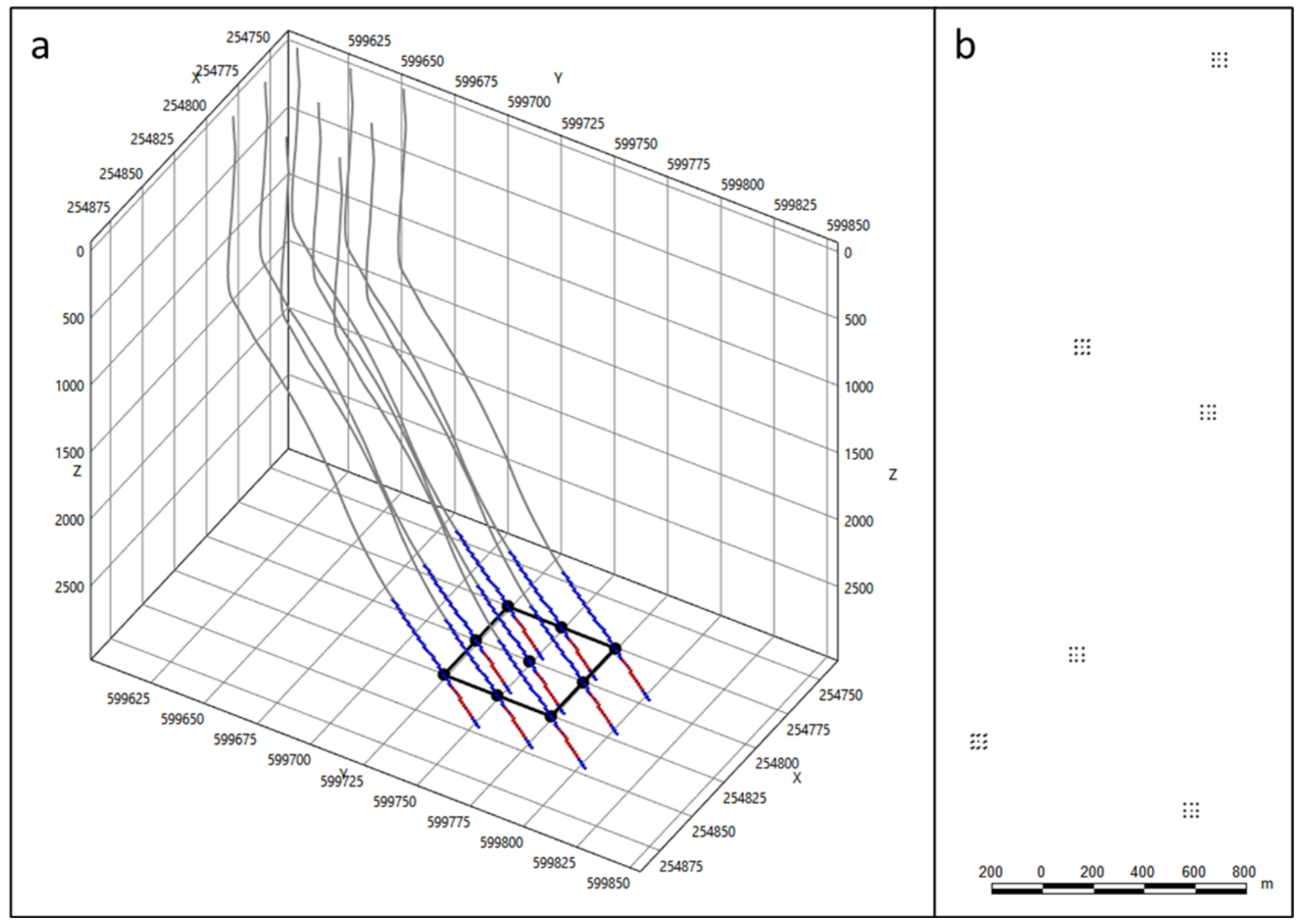

Figure 4. On the left (a) is an example of creating copies of one well in 3D projection. The main and created well trajectories are shown. The colours on the trajectories show the classes (blue - 'dry', red - 'gas reservoir'). On the right (b), top view of the intersection of these trajectories with the reference surface.

Using the sgz0 augmentation method, the training sample was increased nine times. To evaluate the effectiveness of this method, further training will be conducted separately on two samples - base and augmented (sgz0) (Table 2).

Table 2. Number of class labels for the base dataset and the augmented dataset

| class | base | sgz0 |
| --- | --- | --- |
| 0 | 47 243 | 425 199 |
| 1 | 6 215 | 55 923 |
| sum | 53 458 | 481 122 |

## 5.4. Test dataset

The main purpose of the formation and use of a test sample is to assess the quality of the training model and the accuracy of forecasting.

To achieve this goal, after creating a training dataset, 42 wells were excluded from it. The generated test dataset was excluded from the technological stack already at the stage of feature selection and throughout the entire process of model training, up to the evaluation of the effectiveness of the meta-model. This was done to prevent data leakage from the training set to the test set, which made it possible to ensure the correctness of the assessment of the effectiveness of the gas saturation forecast.

The selection of test wells was carried out randomly, one well from the cluster, and from the list of exploration and exploratory wells so that the total number of wells was about 10% of the total number of wells (425) in the study, and the class ratio was 0.13 in total in both the test and training samples. The spatial position of the test wells is shown by red crosses with the name of the wells in Figure 1.

## 6. Machine learning design

The work used a technological stack of several methods for combining machine learning algorithms. At the first stage, the bagging method was used, when an ensemble of base models was created, which were trained on bootstrap samples [9]. Some of the base models themselves were ensembles of decision trees with gradient boosting. At the second stage, the stacking method was used for the final forecast [10] (Fig. 5).

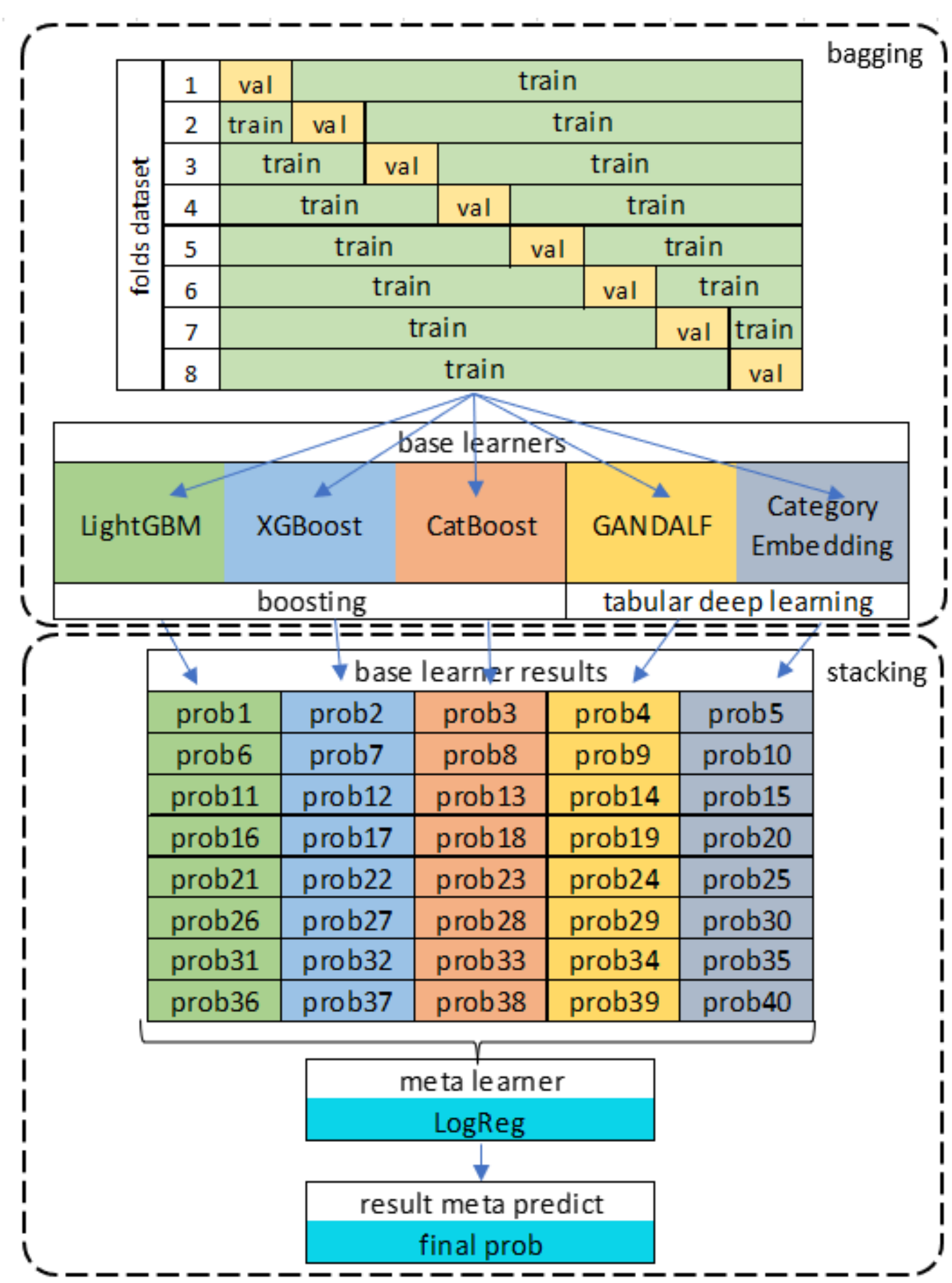

Figure 5. Overall scheme of the process stack for classifier training and probability prediction

### 6.1. Bagging

Bootstrap Aggregating creates a subset of models-base algorithms based on bootstrap samples from the original dataset. The samples were created based on data from 383 wells. The wells were randomly divided into eight parts with the same fixed list of wells for bootstrap samples. In each variant, algorithms were trained at ~335 wells, and validation at ~48. For the augmentation option, copies of wells with new vectors from attributes were added to the wells. Thus, each base model viewed the data eight times from different "points of view" (Fig. 5 bagging part).

## 6.2. Stacking

The integration and generalization of the knowledge gained from each individual model was carried out through ensembling using the stacking method. This method involves using the outputs of several base models (the first level) as input data for the meta-model.

The choice of base models in stacking plays a key role, since the effectiveness of the ensemble depends on it. For successful application of the stacking method, it is recommended to use strong models that have different approaches and algorithms. This diversity helps to improve the generalizing ability of the ensemble, as different models can capture different aspects of the data and identify unique patterns by making mistakes in different places.

The meta-model combines the knowledge of the base models, providing a better final prediction (Fig. 5, stacking part).

## 6.3. Setting up hyperparameters

Hyperparameters were configured for all algorithms using the Optuna search optimizer library [12] and the tree-structured Partzen estimator (TPESampler) [13, 14].

Optimization was carried out by maximizing the Matthews correlation coefficient (MCC) on the validation part of the sample [15]. MCC is a statistical measure used to assess the quality of binary classification. In the context of this work, it is an optimal metric because it takes into account all four categories of results: true positive, true negative, false positive and false negative. This allows us to get a more complete picture of the performance of the model [16], especially in conditions of class imbalance (Table 2).

## 6.4. Selection of base algorithms for machine learning

The choice of base classification algorithms was carried out on the basis of a single bootstrap sample, followed by optimization of hyperparameters for one hour. The effectiveness of the algorithms was evaluated on a deferred test dataset.

The paper uses three implementations of the algorithm based on gradient boosting over decision trees: CatBoost [17], LightGBM [3] and XGBoost [18]. Four algorithms with a deep learning architecture for working with tabular data implemented in the PyTorch Tabular library [19]: GANDLF [20], Category Embedding [19], FT Transformer [21], TabNetModel [22].

Table 3. Comparative efficiency of algorithms

| **models** | **mcc** |
|---|---|
| LightGBM | 0.74488 |
| XGBoost | 0.74165 |
| CatBoost | 0.73979 |
| Category Embedding | 0.72719 |
| GANDALF | 0.72389 |
| FT Transformer | 0.71197 |
| TabNet | 0.69379 |

All models demonstrate high MCC values, which indicates their ability to classify with acceptable accuracy (Table 3).

For further work, algorithms based on boosting decision trees were selected as the most effective for this dataset. In addition, two deep learning algorithms were applied — Category Embedding and GANDLF — in order to increase the diversity of the final ensemble of classifiers.

## 6.5. Training of base models

After selecting hyperparameters, the models were trained on a training dataset with retrain control on a validation sample. Then the obtained models were further trained on the combined train and validation dataset, adding a certain number of iterations equal to 20% of the value of the best iteration of the initial model training on validation (Fig. 6).

Thus, 40 models were trained for the base and augmented datasets (Fig. 5). For each base model, the importance of features was assessed using the Shapely index and the quality of the forecast was assessed on a test sample.

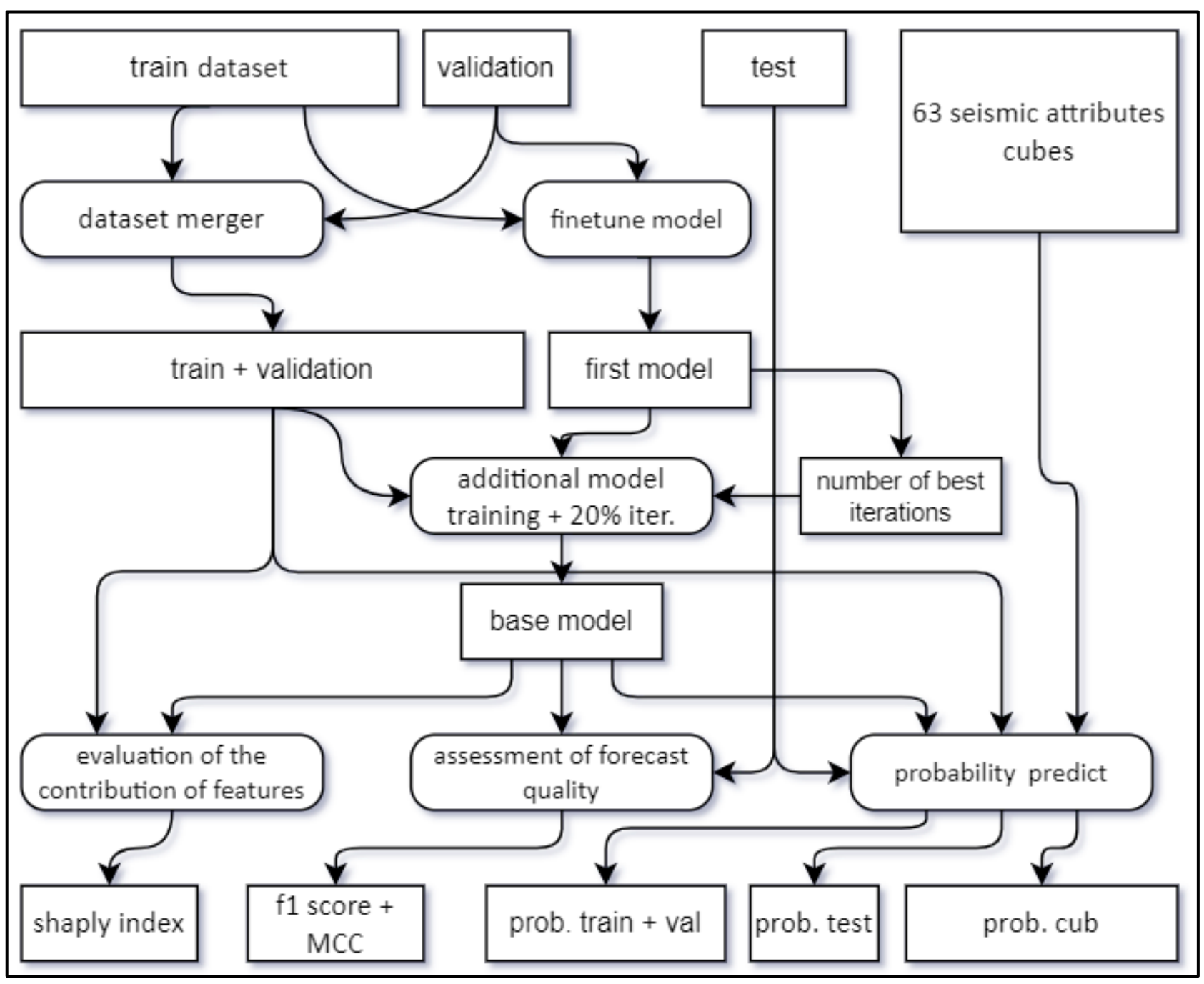

Figure 6. Scheme for training, estimation and prediction of the base model

## 6.6. Training meta-model

The logistic regression algorithm is used as a meta-model (Fig. 5). This is one of the most common and effective machine learning methods for solving classification problems. Logistic regression allows you to predict the probability of an object belonging to one of two classes based on the values of its features. Provides interpretable results in the form of coefficients characterizing the effect of each feature on the probability of assigning an object to a particular class. The regularization parameters of logistic regression were selected using the TPESampler optimizer by maximizing the average MCC metric on cross-validation with 5 folds.

# 7. Assessment of models and results

## 7.1. Assessing the quality of classifier ensemble prediction on the baseline and augmented datasets

At the first stage, as a result of a comparative analysis of two models based on key classification metrics (variations of the f1-score and MCC metrics), the advantage of a model trained on augmented data was determined. For Class 1, the sgz0 model showed an f1 score of 0.7949 compared to 0.7268 for the base model. The Matthews correlation coefficient for sgz0 was 0.7689, which is higher than the base model and indicates a better and more balanced classification (Table 4).

Table 4. Prediction quality metrics of the base and augmented models

| metrics | base | sgz0 | support |
|---|---|---|---|
| f1 class 0 | 0.9677 | 0.9733 | 4 530 |
| f1 class 1 | 0.7268 | 0.7949 | 614 |
| accuracy | 0.9423 | 0.9528 | 5 144 |
| f1 macro avg | 0.8472 | 0.8841 | |
| f1 weight. avg | 0.9390 | 0.9520 | |
| MCC | 0.7025 | 0.7689 | |

At the second stage, the McNemar test was conducted, which is used to evaluate the effectiveness of binary classification algorithms. This statistical method allows you to determine whether there are significant differences in prediction accuracy between two classifiers on the same set of test data.

The test is especially useful in situations where the overall accuracy of the models may be similar, but they make mistakes on different samples. By analyzing the differences in predictions, the McNemar test helps to identify whether one classifier is actually superior to another in a statistically significant way.

The null hypothesis of the McNemar test states that there is no statistically significant difference in performance between the two classifiers being compared.

Table 5. McNemar's contingency table

| | sgz0 correct | sgz0 wrong |
|---|---|---|
| base correct | 4821 | 26 |
| base wrong | 64 | 233 |

To test the null hypothesis that the predictive power of the two models is equal (using a significance level of 0.05), the McNemar test was performed (Table 5).

As a result, chi-squared criterion values of 15.21 and a p-value of 0.000096 were obtained. As the resulting p-value is significantly less than the significance level, the null hypothesis is rejected - there is a statistically significant difference in the strength of prediction between the models.

Considering that the sgz0 model made 2.46 times fewer errors than the base model (Table 5), and that the classification metrics are better (Table 4), the model trained on the augmented sgz0 data is a more effective tool for solving the classification task than the model trained on the base dataset.

The absolute values of the prediction quality metrics for the sgz0 model indicate its acceptable reliability in solving the class separation problem based on well data on rocks with different gas saturation and predictors – attributes of the seismic wave field. This model was used for subsequent analysis and final prediction.

### 7.2. Assessing the importance of base classifiers

The contribution of the base models trained on different bootstrap samples to the final prediction of the meta-model was evaluated.

The contribution of each model was determined by calculating the Shapley index for each variant, after which the index value was normalized by the total sum of the values. The sum was then calculated according to algorithms, resulting in the sum of the contribution of all options being equal to one (Table 6).

Table 6 shows the results of the evaluation of the machine learning algorithms, including their significance and prediction quality indicators measured by MCC.

Table 6. Contribution by importance to the final prediction of the algorithms and aggregated MCC values

| rank | algoritm | importance | mcc max | mcc mean | mcc std |
|---|---|---|---|---|---|
| 1 | LightGBM | 0.234 | 0.7573 | 0.7469 | 0.0076 |
| 2 | XGBoost | 0.223 | 0.7532 | 0.7438 | 0.0067 |
| 3 | Category Embedding | 0.195 | 0.7461 | 0.7352 | 0.0062 |
| 4 | CatBoost | 0.178 | 0.7410 | 0.7370 | 0.0047 |
| 5 | GANDALF | 0.170 | 0.7398 | 0.7306 | 0.0059 |

The LightGBM algorithm shows the greatest cumulative significance in terms of contribution to the prediction of the meta-model, amounting to 0.234, among all the algorithms considered. In addition, it shows the maximum MCC values on the base models.

XGBoost ranks second in terms of importance for the final prediction with a score of 0.223. The Category Embedding deep learning algorithm outperformed the CatBoost tabular data SOTA algorithm in terms of maximum MCC values and importance.

However, in general, the impact of each algorithm on the final prediction is significant, and the contribution of each algorithm to the final model is relatively uniform.

At the same time, the variety of models in the ensemble is a key factor in ensuring higher performance compared to using individual models.

Integrating different but powerful models into an ensemble allows the ensemble to process different types of data and patterns more efficiently. Each model may have its own unique advantages and may be able to identify certain data features that may be missed by other models.

The final ensemble, obtained using the method of stacking predictions of base models by the logistic regression algorithm, demonstrated an efficiency of MCC 0.7689, which exceeded the maximum values of individual best models.

### 7.3. Assessing the importance of attributes

For each of the base models, the significance of the features required for classification was assessed. This evaluation was performed using the Shapley index. The values obtained were multiplied by the importance factor of the model in the ensemble. The values for each feature were then summed and normalized so that the sum of the values for all 63 features was 100%.

As a result, the ten most significant features for seismic event prediction were determined, as well as their tuning parameters (Table 7).

Table 7. Ten most important seismic attributes for effective classification and their tuning parameters

| rank | imp | algorithms | parameters |
|---|---|---|---|
| 1 | 7.8% | reflection intensity | window_size=40 |
| 2 | 4.1% | reflection intensity | window_size=80 |
| 3 | 3.3% | rms | window_size=80 |
| 4 | 3.2% | coherence | inline_window_size=36, crossline_window_size=36, vertical_window_size=64 |
| 5 | 2.8% | rms | window_size=20 |
| 6 | 2.7% | chaos | sigma_x=1.5, sigma_y=1.5, sigma_z=10 |
| 7 | 2.5% | coherence | inline_window_size=36, crossline_window_size=18, vertical_window_size=64 |
| 8 | 2.5% | amplitude spectrum | |
| 9 | 2.3% | spectral decomposition | wavelet_type=morlet, channel_frequency=35, length=200, frequency=50 |
| 10 | 2.3% | variance | inline_window_size=10, crossline_window_size=10, vertical_window_size=60 |

Reflectance intensity is the most significant parameter, which is the top-ranked parameter twice with different window sizes. The RMS (root mean square) of the amplitudes is also an important attribute that varies with different window sizes. Coherence, another important parameter, is also mentioned twice with different parameter configurations.

Although these attributes clearly dominate in terms of importance, the remaining attributes collectively contribute significantly to the overall prediction system.

The analysis of the top ten attributes for prediction shows the importance of selecting and tuning seismic attribute generation algorithms to improve the performance of machine learning algorithms.

## 8. Gas saturation prediction and approximation

After evaluating the prediction quality, the sgz0 meta-model was trained on all available data, then the model was calibrated on isotonic regression. After calibration, prediction was performed. A three-dimensional cube of calibrated probabilities of the study space belonging to class 1 "gas reservoir" was obtained.

To demonstrate the 3D probability cube, its 2D approximation - a map of gas thicknesses in the studied interval - was calculated. The predicted gas thicknesses were calculated using the following formula:

$$h_{ij} = \sum_{k=1}^{K} 1_{\left(P_{ijk} \geq P_T\right)}\, s \quad ,$$

where $s$ vertical resolution of the approximation grid, $P_{ijk}$ predicted calibrated probability of voxel belonging to the "gas reservoir" class, $P_T$ threshold value of probability of voxel belonging to the "gas reservoir" class, $1_{(\cdot)}$ indicator function.

A vertical approximate grid resolution of 8 meters per voxel was used to create the map. The probability threshold was set to 0.5. This means that each pixel of the map is color coded to represent the sum of voxels whose probability threshold exceeded 0.5, with a resolution of 50 by 50 meters (Figs 7a, 7b).

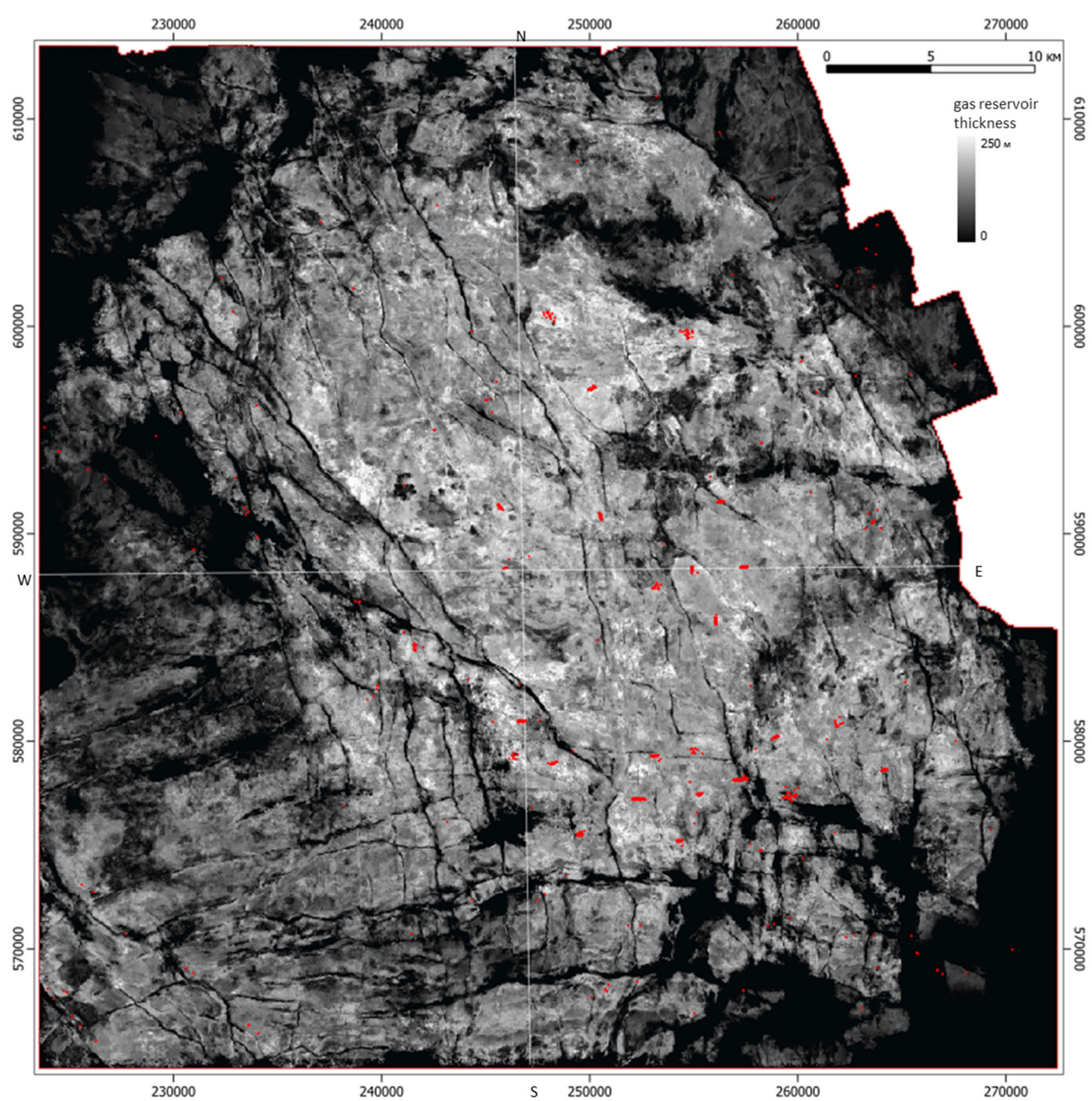

Figure 7a. Final predictive map of gas reservoir thickness. The red line is the study area. Red dots - wells of the training data set. White lines - seismic cross-sections with probability of gas reservoirs. Well positions - intersection of well trajectories with the plane at 2000m

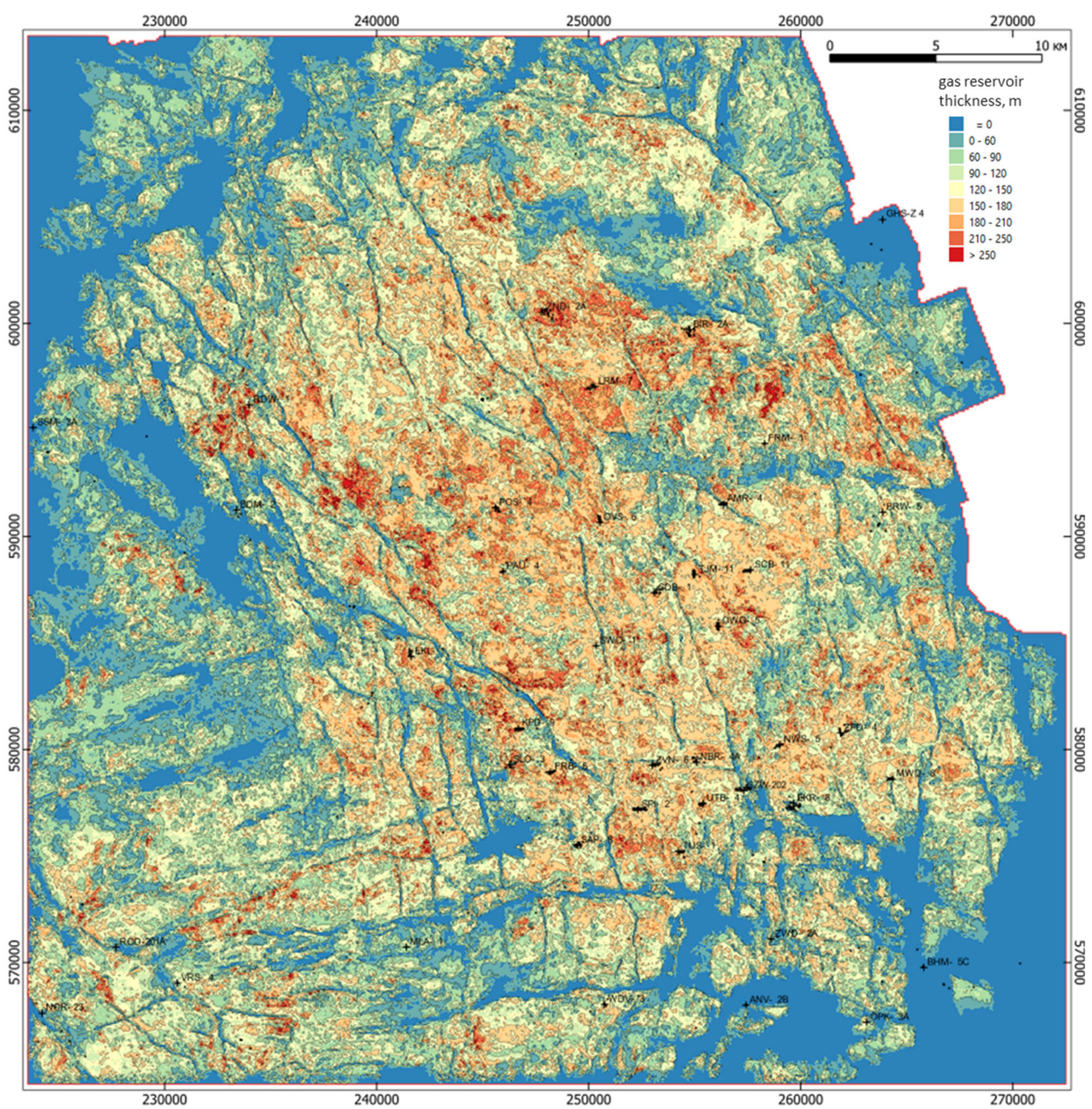

Figure 7b. Final predicted gas reservoir thickness map in conventional coloring. The red line is the study area. Black dots are wells in the training dataset. Black crosses with well names - test wells. Well positions - intersection of well trajectories with the plane at 2000m.

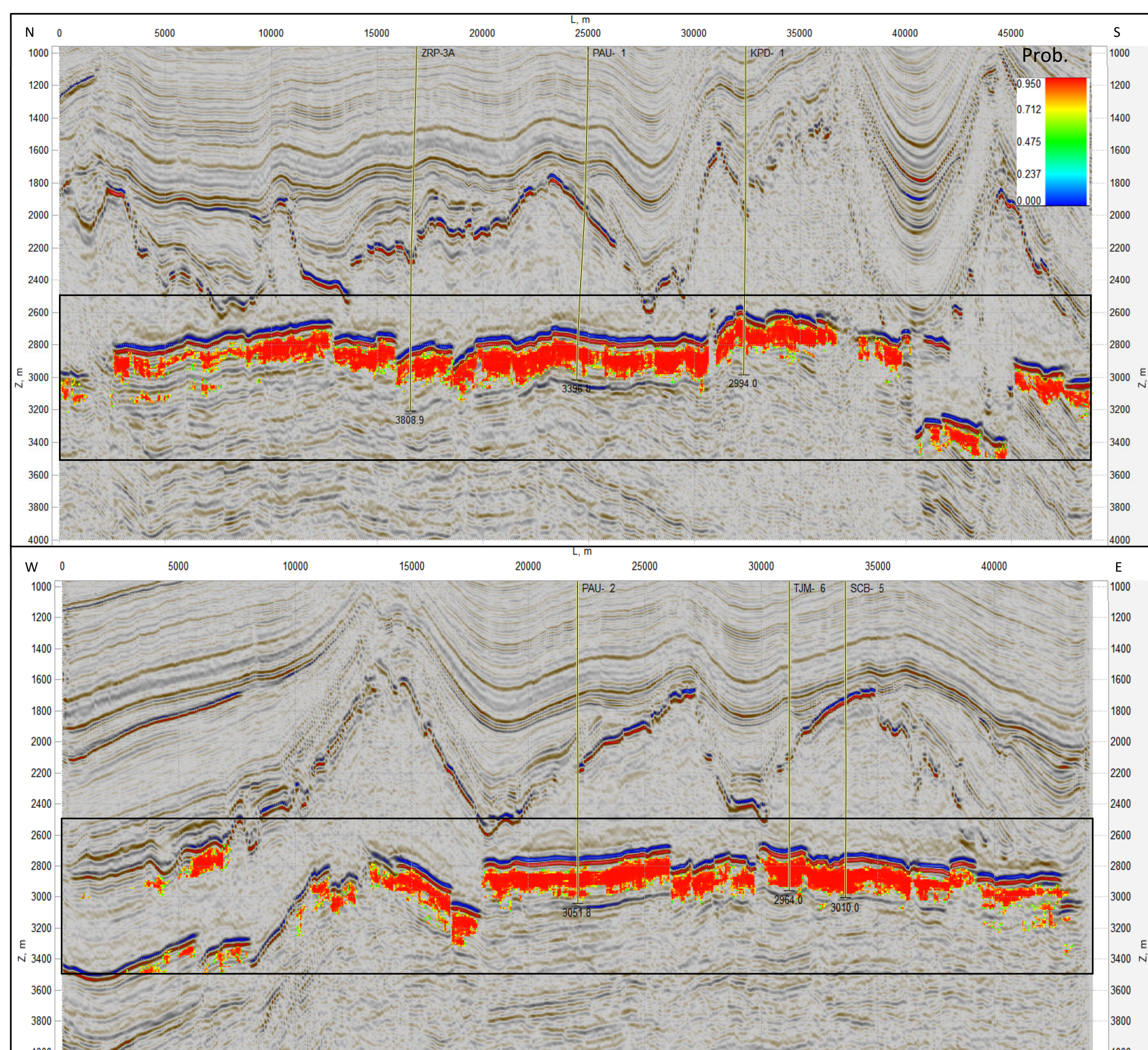

Figure 8. Seismic cross-sections with probability of gas reservoirs. The color palette encodes the probability of belonging to the "gas reservoir" class. Probabilities below 0.5 inclusive are made transparent. The black frame highlights the interval on the probability prediction slice.

## 9. Conclusions

This paper presents a proof of concept for spatial prediction of rock saturation probability using classifier ensemble methods. Two types of information were used as input data for the prediction: extracted classes from saturation logs on 425 wells and one seismic cube.

In the study, 1481 seismic attributes of the seismic field were generated, from which 63 of the most important ones for the predictive algorithms were selected using the proposed feature selection methods. The degree of importance of the used attributes and classification algorithms for the final ensemble model was demonstrated. The efficiency of the proposed method of well and seismic data augmentation – one step grid with zero vertical offset (sgz0) – with relative to the base dataset was shown.

An ensemble classifier based on the augmented data, using 63 seismic field attributes and extracted classes from gas saturation logs, was used to predict rock gas saturation probabilities (Figs. 7a, 7b, 8). The results of evaluating the final metamodel on a test sample of 42 wells (blind well test) demonstrate the effectiveness of the prediction: the Matthews correlation coefficient is 0.7689 and the F1-score for the gas reservoir class is 0.7949. These values are high in a wild using a single seismic cube with a single processing graph. The final meta-model should exceed these values because the meta-model was pre-trained on all available data after evaluating the classification performance.

After approximation of the 3D probability space (Fig. 8) to the map, the gas thickness map clearly shows a geologic pattern – a system of extended tectonic faults that play a key role in controlling gas saturation (Fig. 7a). Practice shows that the quality of the prediction can be improved, for example, by adding seismic cubes with different processing schedules and more complete interpretation of well logs to the process stack.

The proposed approach has a certain independence from expert opinions, each of its stages is formalized and aimed at optimizing the classification efficiency metrics. Thus, this approach can become a primary tool for generalization of available geological and geophysical information for further work of geologists and geophysicists. The approach allows to create a three-dimensional probabilistic representation of a geological object, improve it taking into account new data and carry out a continuous cycle of formalized assessments.